\documentclass[bibyear]{aa} % if the references are not structured 
\usepackage{graphicx}
\usepackage{txfonts}
\usepackage{natbib}
\usepackage{xfrac}
\usepackage{color}           % For color text: \color command
\newcommand{\R}[1]{{#1}} %\newcommand{\R}[1]{{\color{red}{\underline{#1}}}} 

\begin{document} 

\title{Thermodynamics of supra-arcade downflows in solar flares}

\author{Xin Chen\inst{1}, Rui Liu\inst{2,3}, Na Deng\inst{1} \and Haimin Wang\inst{1}}

\authorrunning{X. Chen, R. Liu, N. Deng \& H. Wang}

\institute{Space Weather Research Laboratory, New Jersey Institute of Technology, University Heights, Newark, NJ 07102-1982, USA\\ 
\email{xc55@njit.edu}
\and
CAS Key Laboratory of Geospace Environment, Department of Geophysics and Planetary Sciences, University of Science and Technology of China, Hefei, Anhui 230026, China\\ 
\and 
Collaborative Innovation Center of Astronautical Science and Technology, Hefei, Anhui 230026, China\\}

\date{Received ... ; accepted ...}

% \abstract{}{}{}{}{} 
% 5 {} token are mandatory
 
  \abstract
   {Supra-arcade downflows (SADs) have been frequently observed during the gradual phase of solar flares near the limb. In coronal emission lines sensitive to flaring plasmas, they appear as tadpole-like dark voids against the diffuse fan-shaped ``haze'' above, flowing toward the well-defined flare arcade.}
   {We aim to investigate the evolution of SADs' thermal properties, and to shed light on the formation mechanism and physical processes of SADs.}
   {We carefully studied several selected SADs from two flare events and calculated their differential emission measures (DEMs) as well as DEM-weighted temperatures using data obtained by the Atmospheric Imaging Assembly (AIA) onboard the Solar Dynamic Observatory. }
   {Our analysis shows that SADs are associated with a substantial decrease in DEM above 4 MK, which is 1--3 orders of magnitude smaller than the surrounding haze as well as the region before or after the passage of SADs, but comparable to the quiet corona. There is no evidence for the presence of the SAD-associated hot plasma ($>20$ MK) in the AIA data, and this decrease in DEM does not cause any significant change in the DEM distribution as well as the DEM-weighted temperature, which supports this idea that SADs are density depletion. This depression in DEM rapidly recovers in the wake of the SADs studied, generally within a few minutes, suggesting that they are discrete features. In addition, we found that SADs in one event are spatio-temporally associated with the successive formation of post-flare loops along the flare arcade.}
   {}


   \maketitle

%
%-------------------------------------------------------------------
\section{Introduction}

Solar flares are among the most energetic events in the solar system. Despite years of study, some processes and features in flares remain poorly understood; among them are supra-arcade downflows (SADs; \citealt{McKenzie1999,Innes2003,Asai2004}). SADs are tadpole-like dark voids falling sunward through the fan-shaped, haze-like flare plasma above the flare arcade (also referred to as ``supra-arcade fan'' in the literature) in soft X-rays and extreme ultra-violet (EUV). Furthermore, SADs are generally considered as a signature of the outflow of magnetic reconnections occurring high in the corona. Their presence is found to be co-temporal with hard X-ray and microwave bursts, which implies their close relation to the impulsive release of magnetic free energy \citep{Asai2004,Khan2007}.

Even though SADs are not likely to be rare in flares, their detection requires specific conditions. As a faint structure, SADs are preferentially seen against a bright background emission (generally the haze region). Further, they could be obscured by bright structures in the foreground due to the optically thin nature of soft X-rays or EUV in the corona. Therefore, the observations of SADs are almost exclusive to the limb flares \citep{Savage2011}, whose flare arcade is oriented in the North-South direction, so that the axis of the arcade is perpendicular to the line-of-sight (LOS).

Numerous samples of SADs have been studied for their kinematics in the last decade. Well above the flare arcade, SAD's speed ranges from several dozens to a few hundred kilometers per second without significant acceleration or deceleration, but the most complete flow paths show significant deceleration approaching the flare arcade \citep{McKenzie2009,Savage2011}, just like downward shrinking loops \citep[see e.g., Fig3 in][]{Zhu2016}. As more cases have been studied using various instruments, the interpretations of SADs are also evolving. Early on SADs were assumed to be the cross-sections of evacuated magnetic flux tubes \citep{McKenzie2000,McKenzie2009}. Recently, \citet{Savage2012} reinterpreted SADs as the wake, that is, the trailing region of retracting magnetic flux tubes. Although this may explain some observations from certain aspects, the physics involved in this reinterpretation remains obscure \citep{Scott2013}. More recently, \citet{Liu2013} suggested that a SAD is a twisted mini-flux rope that is highly stretched above the cusp region before it finally shrinks and untwists to become a flare loop.

The plasma properties of SADs, which are critical in the exploration of their physical mechanism, become more feasible using models or derivatives from observations in recent years. Previously, SADs were conjectured as evacuated flux tubes, which implies that they would be less dense but hotter than the surrounding plasma \citep{McKenzie1999}. Recently, through magneto-hydrodynamics (MHD) modeling, \citet{Maglione2011} and \citet{Cecere2012} simulated a similar sunward void structure caused by the interaction of shocks with expansive waves during the reconnection. In their models, these SAD-like voids are of low densities, high beta values, and high temperatures that could reach more than 80 MK \citep{Maglione2011} or 20-30 MK \citep{Cecere2012}. On the other hand, based on the high-cadence multi-wavelength observations provided by the Atmospheric Imaging Assembly (AIA) onboard the recently launched Solar Dynamic Observatory (SDO), the plasma density and temperature can be recovered using the differential emission measure (DEM) method. The emission measure of observed SADs is lower than the surrounding plasma, as expected for a density depletion \citep{Savage2012,Hanneman2014}. Further, \citet{Hanneman2014} found that SADs are generally cooler than the surrounding plasma at around 10 MK. In addition to SDO/AIA, the authors included \textit{Hinode}/X-Ray Telescope which is sensitive to a hotter temperature range (10-100 MK) but there is still no evidence of an extremely hot component exceeding 20 MK present in the DEM of SADs. 

The mechanism of SADs has been under active investigation, especially via modeling and simulation in recent years. \citet{Scott2013} supported the interpretation of SAD as a flux tube embedded within the current sheet descending toward the solar surface \citep{Savage2012}. Their simulation shows that a region of depletion follows the shock in front of the descending flux tube. \citet{Cassak2013} suggested that the reconnection is temporally continuous, resulting in sustained void channels that would not be filled by surrounding high density plasma. However, it is noted that the occurrence height of these void channels and the preferred LOS to detect them, which is parallel to the axis of a flare arcade, are inconsistent with the observations of SADs \citep{Guo2014}. Some SAD observations show a leading retracting loop \citep[e.g.][]{Savage2012}, but others apparently follow neither emission enhancement nor retracting loops \citep{Innes2014}. Hence it is still an open question whether a successful model of SAD should include the retracting flux tube as a necessary component. More recently, \citet{Guo2014} demonstrated that Rayleigh-Taylor instabilities in the downstream region of the reconnecting current sheet are a promising mechanism of SADs. Their simulated AIA 131 \r{A} data compare favorably with the SADs observed between spike-like bright plasma \citep{Innes2014}. Meanwhile, \citet{Cecere2015} suggested that a turbulent current sheet resulting from a combination of the tearing mode and Kelvin-Helmholtz instabilities would play an important role in producing SADs. 

In this paper, we analyze several well-observed SADs from two flare events using SDO/AIA. A newly developed DEM method \citep{Hannah2012,Hannah2013} is employed to diagnose their thermal properties. In regards to the DEM-weighted temperature, we found that the DEM contributed by SAD itself is negligible, which supports this idea that SADs are density depletion. Instead of following the motion of an individual SAD through the corona, we took an Eulerian approach to investigate how SADs impact on their environments, aiming to shed new light on their nature and to provide useful constraints for models. The instrumentation and DEM method are introduced in Section~\ref{sect:obs}, observations and analysis are presented in Section~\ref{sect:res}, and concluding remarks are given in Section~\ref{sect:dis}.

%-------------------------------------------------------------------
\section{Observations and Method}
\label{sect:obs}

\subsection{Observation} 
\label{subsect:obs}

This study is based on the data obtained by AIA \citep{Lemen2012} onboard SDO \citep{Pesnell2012}. AIA contains ten EUV and UV channels covering a full-disk field of view (FOV; up to $\sim$1.3 $R_{\odot}$) with a spatial resolution of $\sim$1$''$s. The six EUV channels (94 \r{A} , 131 \r{A}, 171 \r{A}, 193 \r{A}, 211 \r{A}, 335 \r{A}), taking images at a high cadence of 12 sec, are sensitive to a variety of coronal temperatures (from $10^{5.5}$ K to $10^{7.5}$ K). We processed the AIA level-1.0 data with the SolarSoftWare (SSW) procedures, \verb+aia_deconvolve_richardsonlucy.pro+ and \verb+aia_prep.pro+, to produce level 1.6 data. The Point Spread Function for deconvolution is calculated by \verb+aia_calc_psf.pro+.

We investigated two flares in this study and one of them, the 2013 October 2 event, is coincident with a scheduled calibration of the Helioseismic and Magnetic Imager onboard SDO. For AIA images, this calibration caused bad frames lasting about 2 minutes every 12 minutes due to the rotation of the telescope. The cadence of AIA EUV channels is also extended to 20 seconds during the calibration. Except for those bad frames, the AIA image quality is normal and stable. Thus, for this particular event, we performed an additional process that removed all the bad frames and carefully aligned the limb position for each channel. There is no such complication in the other event, the 2014 Apr 2 flare. We use the most recent available AIA temperature response functions (V4 calibration), applying both the CHIANTI fix \footnote{http://sohowww.nascom.nasa.gov/solarsoft/sdo/aia/response/chiantifix\textunderscore notes.txt} to account for missing emission lines in the CHIANTI database in channels 94 and 131 \r{A} as well as EVE normalization to give good agreement with full-disk EVE spectral irradiance data.
 
\subsection{Differential emission measure} 	
\label{subsect:DEM}

\begin{figure*}
\centering
\includegraphics[width=16cm]{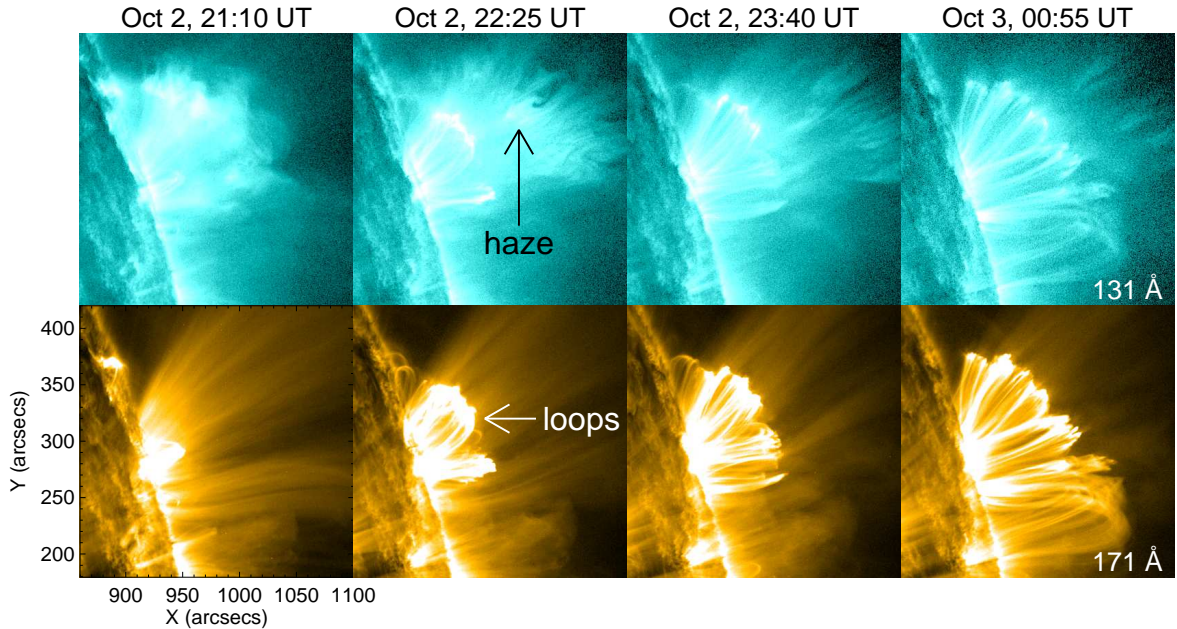}
\caption{Snapshots of AIA images featuring \R{the formation sequence of post-flare loops}. From top to bottom, each row presents multi-wavelengths of AIA images every 75 minutes, which shows that the post-flare loops first appeared in the north side and later expanded southward, most obviously in AIA 171 \r{A}.}
\label{fig311}
\end{figure*}

Generally, in the optically thin solar corona, the emission measure (EM) is proportional to the square of electron density in a certain volume. Accordingly, the differential emission measure (DEM) can be written as

\begin{equation}
DEM(T)= {n_{e}^{2}(h)} \frac{dh}{dT},
\end{equation}

\noindent
where $n_{e}$ is the electron density at temperature $T$ at the position coordinate $h$ along the line-of-sight. The observed signal in the $i$th channel, $g_{i}$, is related with DEM by

\begin{equation}
g_{i}=\int_{T} K_{i}(T)DEM(T)dT + \delta g_{i},
\end{equation}

\noindent
where $K_{i}(T)$ is the temperature response of the $i$th channel and $\delta g_{i}$ is the corresponding measurement error. To recover DEM is a well-known ill-posed inverse problem. One major difficulty of the inversion is to constrain the amplification of measurement error, which has been attempted using several different methods (e.g., \citealt{Weber2004,Aschwanden2011,Kashyap1998}), with their advantages and disadvantages.

The method we used to reconstruct DEM in this work is an enhanced regularization method developed by \citet{Hannah2012}. It is computationally fast and has been employed to analyze several events (e.g., \citealt{Hannah2013,Chen2014,Gou2015}). This code calculated DEM using the six AIA EUV channels. The results yield DEM for each individual pixel, its uncertainty, and the temperature resolution, especially the last one that is rarely provided by other methods. In this study, 12 temperature bins, namely, [0.5-1], [1-1.5], [1.5-2], [2-3], [3-4], [4-6], [6-8], [8-11], [11-14], [14-19], [19-25] \& [25-32] MK, are selected to carry out the DEM analysis. For selected regions of interest (ROIs), we further calculated the average DEM of each region and the corresponding uncertainty, following the rule of error propagation.

%--------------------------------------------------------------------
\section{Result}
\label{sect:res}

\subsection{2013 October 2 event} 
\label{subsect:13Oct2}

\begin{figure*}
\centering
\includegraphics[width=13.5cm]{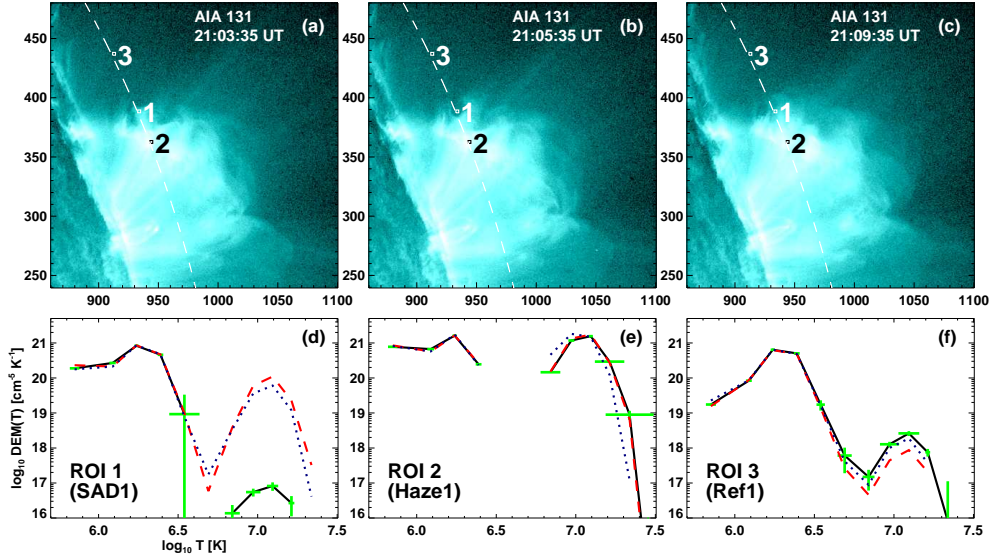}
\caption{DEM analysis for a SAD appearing at around 21:05 UT on 2013 October 2. Panels (a)-(c) show snapshots of AIA 131 \r{A} images before, during, and after the SAD \R{has} passed through ROI1. Panels (d)-(f) show DEM plots for R$_\text{SAD1}$, R$_\text{haze1}$, and R$_\text{ref1}$, respectively (3$''\times$3$''$ each); the red dashed lines correspond to 21:03 UT (panel (a)); the black solid lines correspond to 21:05 UT (panel (b)) with green error bars; and the blue dotted lines correspond to 21:09 UT (panel (c)).}
\label{fig312}
\end{figure*}

\begin{figure*}
\centering
\includegraphics[width=13.5cm]{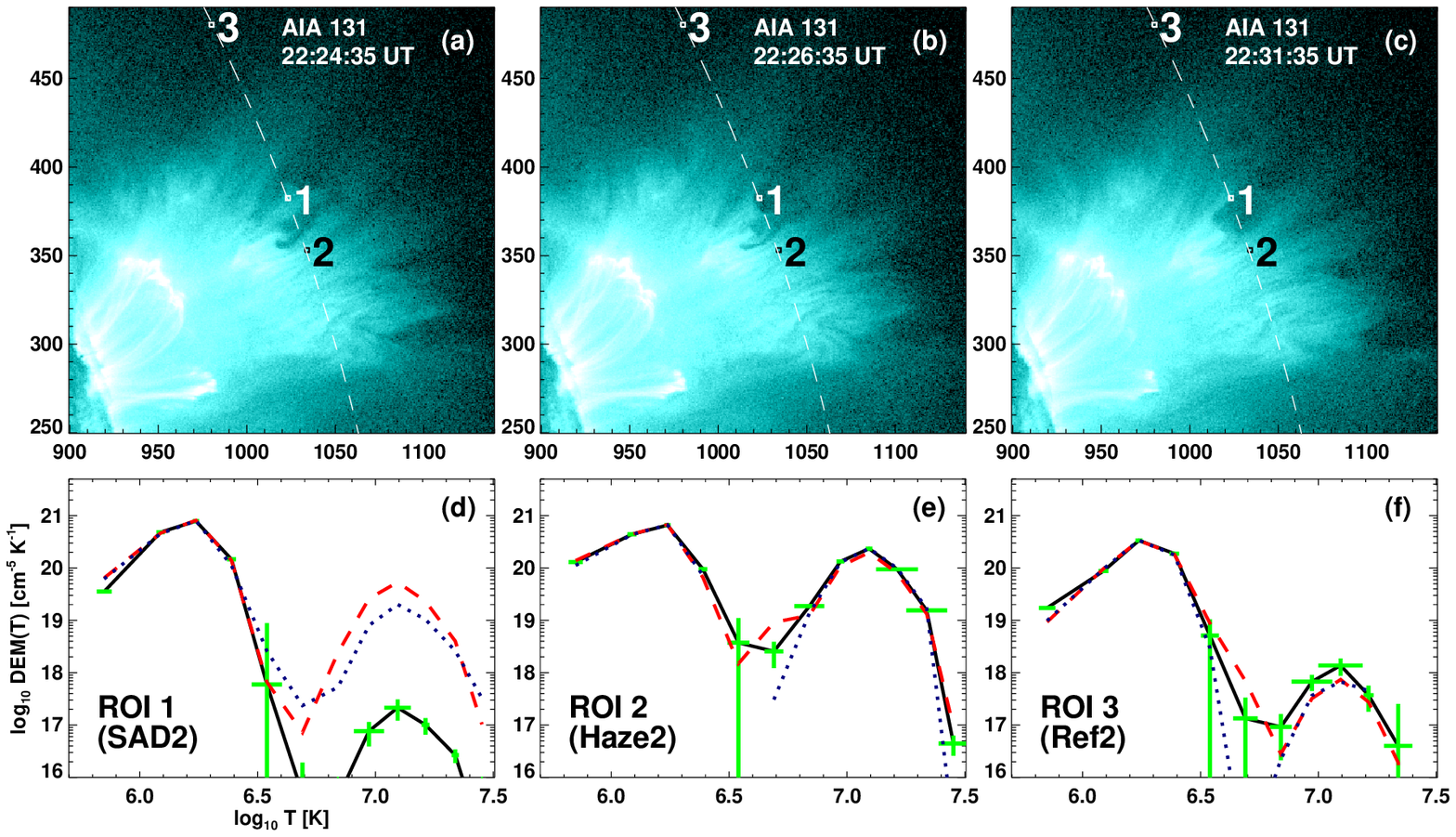}
\caption{DEM analysis for a SAD appearing at around 22:26 UT on 2013 October 2. Panels (a)-(c) show snapshots of AIA 131 \r{A} images before, during, and after the SAD \R{has} passed through ROI1. Panels (d)-(f) show DEM plots for R$_\text{SAD2}$, R$_\text{haze2}$, and R$_\text{ref2}$, respectively (3$''\times$3$''$ each); the red dashed lines correspond to 22:24 UT (panel (a)); the black solid lines correspond to 22:26 UT (panel (b)) with green error bars; and the blue dotted lines correspond to 22:31 UT (panel (c)).}
\label{fig313}
\end{figure*}

\begin{figure*}
\centering
\includegraphics[width=13.5cm]{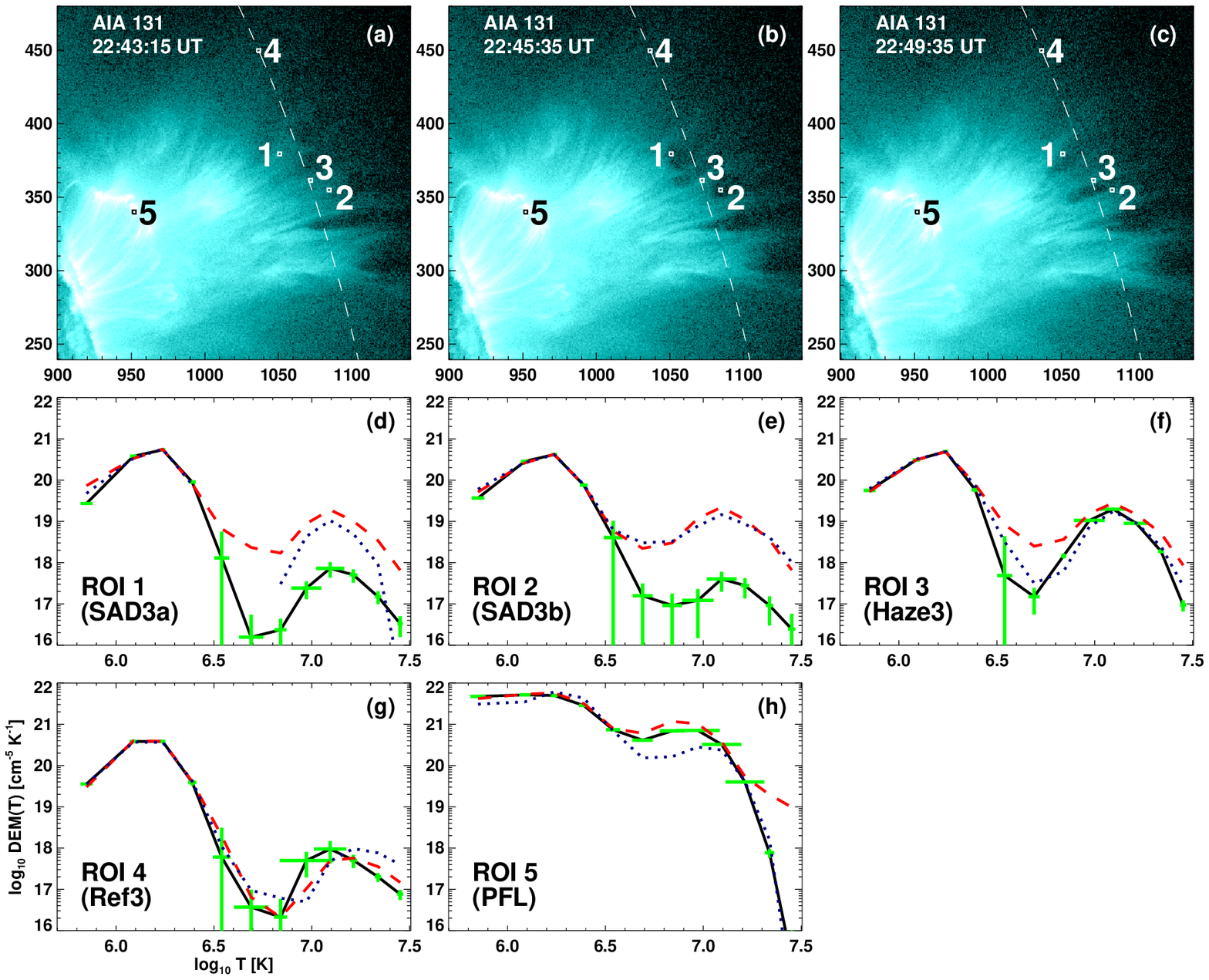}
\caption{DEM analysis for two SADs, 3a and 3b, appearing at around 22:45 UT on 2013 October 2. Panels (a)-(c) show snapshots of AIA 131 \r{A} images before, during, and after the SADs \R{have} passed through ROI 1 and 2. Panels (d)-(h) show DEM plots for R$_\text{SAD3a}$, R$_\text{SAD3b}$, R$_\text{haze3}$, R$_\text{ref3}$, and R$_\text{pfl}$, respectively (3$''\times$3$''$ each); the red dashed lines correspond to 22:43 UT (panel (a)); the black solid lines correspond to 22:45 UT (panel (b)) with green error bars; and the blue dotted lines correspond to 22:49 UT (panel (c)).}
\label{fig314}
\end{figure*}

This flare occurred on the northwestern limb in the NOAA active region (AR) 11850 at around 20:00 UT on 2013 October 2. It is a GOES-class C1 flare partially occulted by the solar limb. The axis of the flare arcade is oriented almost in the longitudinal direction and located very close to the limb. As a result the arcade is observed from a side-on perspective, with a large number of dark SADs flowing through the bright fan-shaped haze from 20:50 UT to 00:50 UT the next day (Oct 3).

In this event, SADs generally exhibit a tadpole shape against the background. When a SAD plunges into the bright haze region, the `head' of the tadpole is apparently compressed into a smaller size. Some SADs do not have a clear wiggling `tail' but an elongated ray behind the falling `head'. These tailing features last several minutes and gradually shrink before disappearing eventually. SADs are seen first above the northern edge of this AR at around N$23\,^{\circ}$ at about 21:00 UT in the AIA 131 \r{A} channel. In the following 2 hours, the locations where they first appear gradually migrate southward (between N$15\,^{\circ}\sim$ N$18\,^{\circ}$, see Figure~\ref{figA1}). Earlier SADs are not only smaller in size, but also travel along shorter and more curved orbits, compared to those observed later. Similarly, we found that the post-flare loops (PFLs) form sequentially in a North-South direction, as shown in Figure~\ref{fig311}. This successive formation of PFLs spans a longer time period ($\sim$4 hours) than the migration of SADs. Only a few PFLs appeared earlier in the low-latitude region.

The DEM method is employed to diagnose the density and temperature of SADs. At around 21:00 UT, the first SAD of interest (SAD$_{1}$) is observed in AIA 131 \r{A} (Figure~\ref{fig312}). We selected a few ROIs for the DEM analysis, including a region on the path of SAD$_{1}$ (R$_\text{SAD1}$), a neighboring region in the haze (R$_\text{haze1}$) and a reference region in the quiet corona at the same height as SAD$_{1}$ (R$_\text{ref1}$). Figure~\ref{fig312}(b) shows a snapshot when SAD$_{1}$ is passing through R$_\text{SAD1}$, as well as shortly before (Figure~\ref{fig312}(a)) and after (Figure~\ref{fig312}(c)) the passage. The corresponding DEM profiles are plotted in Figure~\ref{fig312}(d)-(f). The DEM profile below 4 MK is similar for all ROIs and remains temporally unchanged, which is assumed to account for the coronal background along the LOS. In this study, we focus on the hot ($>$4 MK) DEM component (DEM$_\text{h}$ hereafter) which is associated with flaring plasma and modulated by the passing SADs. For R$_\text{SAD1}$, it is clear that there is a significant drop in DEM$_\text{h}$, by about three orders of magnitude, and then DEM$_\text{h}$ totally recovers soon after SAD$_{1}$ has passed through. When SAD$_{1}$ is located within R$_\text{SAD1}$, DEM$_\text{h}$ is reduced to a value even smaller than that in R$_\text{ref1}$, but the peak temperature of DEM$_\text{h}$ remains unchanged at around 11-14 MK in spite of the drop in DEM$_\text{h}$. In addition, there is no sign of SAD$_{1}$ having a high-temperature component exceeding 20 MK. DEM$_\text{h}$ in R$_\text{SAD1}$ before and after the passage is one order of magnitude smaller than R$_\text{haze1}$, the core of the haze region, probably because R$_\text{SAD1}$ is located at the edge of the haze region. 

Through analyzing the DEM of a few more SADs observed later in this event, we found that generally they share similar characteristics. Figure~\ref{fig313} shows one SAD (SAD$_{2}$) at around 22:26 UT and Figure~\ref{fig314} shows two SADs (SAD$_\text{3a}$ and SAD$_\text{3b}$) at around 22:45 UT. For ROIs with passing SADs (R$_\text{SAD2}$, R$_\text{SAD3a}$, and R$_\text{SAD3b}$), their profiles all present a significant drop when a SAD passes and then fully recover within a few minutes after the SAD {has} left the ROI. DEM$_\text{h}$ in the reference region (R$_\text{ref2}$ and R$_\text{ref3}$) is as weak as those in the ROIs with the presence of SADs. The DEM$_\text{h}$ of haze regions (R$_\text{haze2}$ and R$_\text{haze3}$) is similar to that in the corresponding R$_\text{SAD}$ when the SAD is not inside the region. The DEM$_\text{h}$ still peaks at 11-14 MK for all regions except in the PFL region (R$_\text{pfl}$), whose DEM$_\text{h}$ is apparently shifted leftward, indicating a cooler temperature consistent with the standard picture.

\subsection{2014 April 2 \R{event}} 
\label{subsect:14Apr2}

\begin{figure*}
\centering
\includegraphics[width=13.5cm]{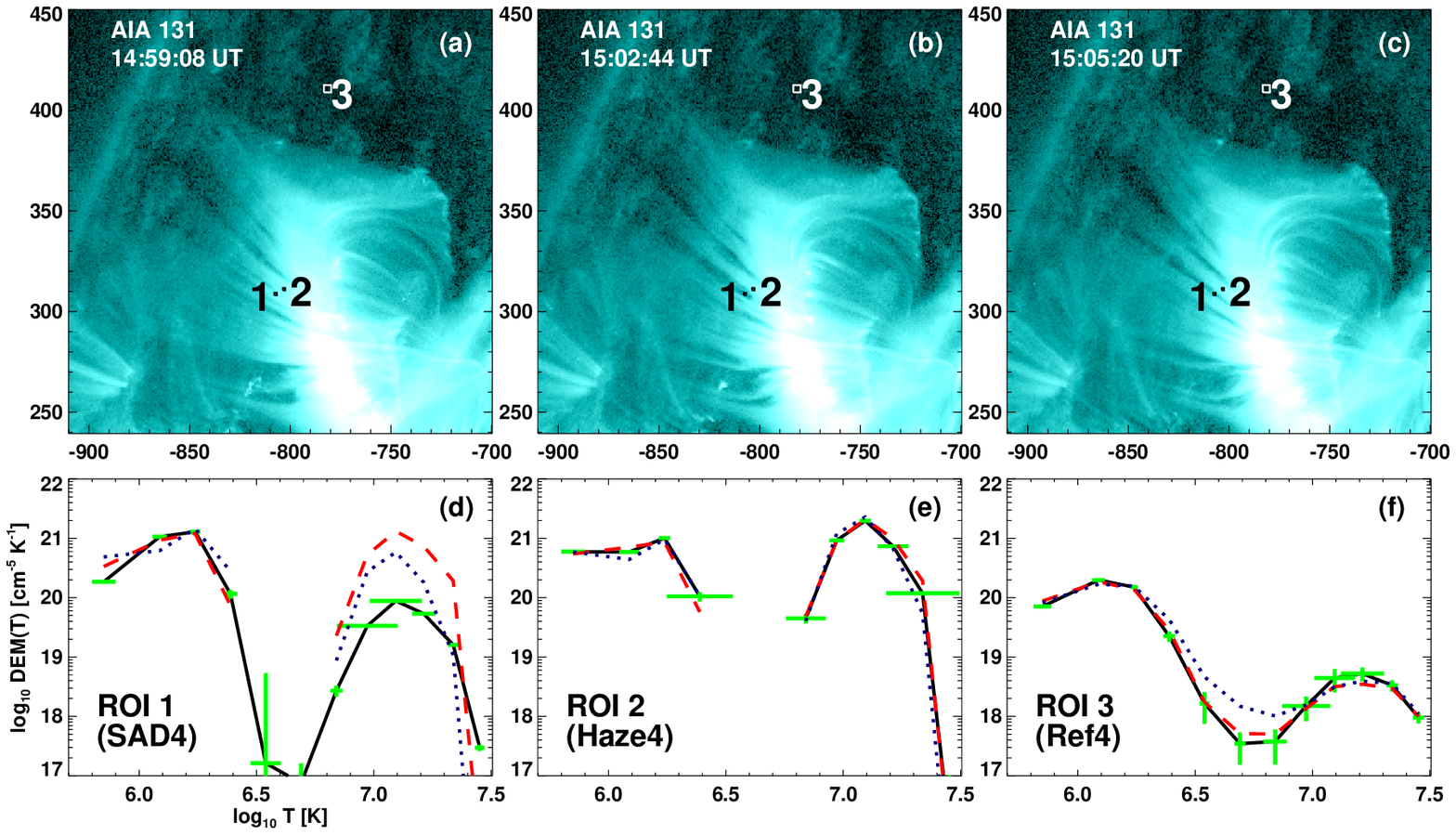}
\caption{DEM analysis for a SAD appearing at around 15:02 UT on 2014 April 2. Panels (a)-(c) show snapshots of AIA 131 \r{A} images before, during, and after the SAD has passed through ROI1. Panels (d)-(f) show DEM plots for R$_\text{SAD4}$ ($1\farcs8\times1\farcs8$), R$_\text{haze4}$ ($1\farcs8\times1\farcs8$), and R$_\text{ref4}$ ($4\farcs2\times4\farcs2$); the red dashed lines correspond to 14:59 UT (panel (a)); the black solid lines correspond to 15:02 UT (panel (b)) with green error bars; and the blue dotted lines correspond to 15:05 UT (panel (c)). }
\label{fig321}
\end{figure*}

\begin{figure*}
\centering
\includegraphics[width=13.5cm]{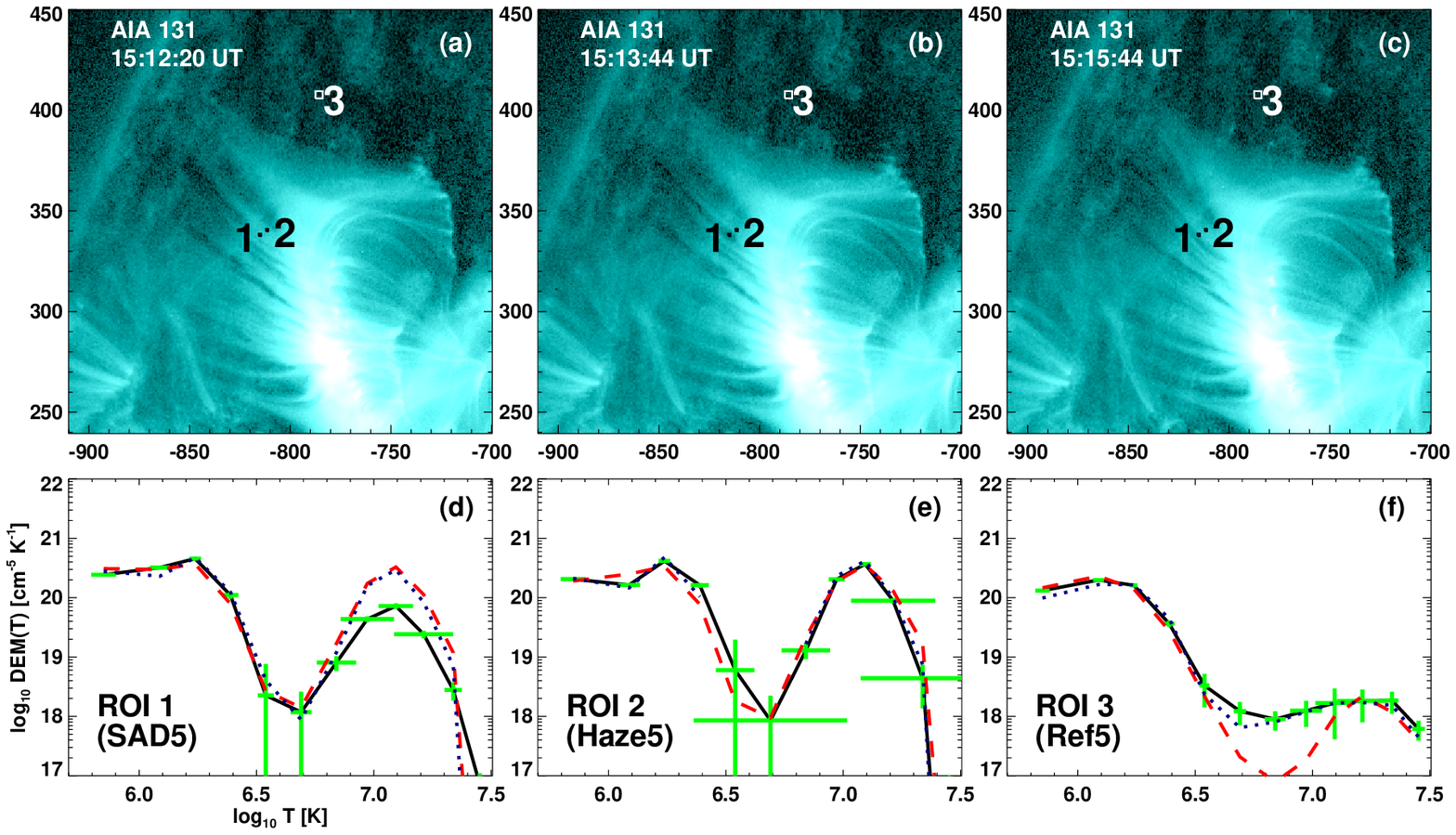}
\caption{DEM analysis for a SAD appearing at around 15:13 UT on 2014 April 2. Panels (a)-(c) show snapshots of AIA 131 \r{A} images before, during, and after the SAD has passed through ROI1. Panels (d)-(f) show DEM plots for R$_\text{SAD5}$ ($1\farcs8\times1\farcs8$), R$_\text{haze5}$ ($1\farcs8\times1\farcs8$), and R$_\text{ref5}$ ($4\farcs2\times4\farcs2$); the red dashed lines correspond to 15:12 UT (panel (a)); the black solid lines correspond to 15:13 UT (panel (b)) with green error bars; and the blue dotted lines correspond to 15:15 UT (panel (c)). }
\label{fig322}
\end{figure*}

This GOES M6.5 flare occurred at around 14:00 UT at N14E53 in NOAA AR 12027. The flare arcade is formed following a filament eruption and has a curved `L' shape. Accordingly, SDO's LOS for this event is oblique to the axis of the flare arcade. Several comet-tail-like bright spikes semi-parallel to one another are located above the top of the flare arcade. In the meantime, SADs are observed to flow between those spikes, most obviously during 14:50 - 16:20 UT.

The SADs in this event still show a tadpole-like shape in AIA 131 \r{A} observations and fade in the bright haze in a few minutes. It is clear in running difference images separated by $\sim$1 min (see Figure~\ref{figA2}) that the reduction in brightness due to the SAD `head' is generally tailed by an enhancement when the plasma recovers to the normal coronal condition. The sizes of SADs appear to be smaller in comparison to the 2013 Oct 2 event probably because of the oblique viewing angle. The supra-arcade region seems to provide stronger contrast in this event, hence almost every SAD is followed by its `tail'. We noted that sometimes the SADs cluster and fall sequentially on a similar path. However, the followers are not necessarily moving along the `tail' of the leaders, considering the LOS confusion and projection effects.
 
The DEM analysis of SADs in this event does not show any major difference from the 2013 Oct 2 event. One case at around 15:02 UT is shown in Figure~\ref{fig321} and another case at around 15:13 UT in Figure~\ref{fig322}. During the passage of SADs, DEM$_\text{h}$ within the ROIs on the path of SADs (R$_\text{SAD4}$ and R$_\text{SAD5}$) are significantly decreased by about one order of magnitude and then recovered soon after that. The magnitude of decrease is smaller than the 2013 Oct 2 event, probably due to the oblique viewing angle in this event, resulting in the presence of hotter and denser haze plasma along the LOS. When SADs are not passing through the aforementioned ROIs, their DEM$_\text{h}$ are very similar to the nearby ROIs (R$_\text{haze4}$ and R$_\text{haze5}$) that are located in the bright spikes. In addition, the peak temperature of all DEM$_\text{h}$ is again more or less stable over time at around 11-14 MK. None of ROIs present any significant DEM beyond 20 MK.

%--------------------------------------------------------------------
\section{Discussion and \R{conclusion}}
\label{sect:dis}

\begin{table*}
\caption{List of $\langle T \rangle_\text{h}$ (in MK) for R$_\text{SAD}$s}
\label{tbl1}
\centering
\begin{tabular}{c c c c c c c}
\hline\hline
Passage of SADs & R$_\text{SAD1}$ & R$_\text{SAD2}$ & R$_\text{SAD3a}$ & R$_\text{SAD3b}$ & R$_\text{SAD4}$ & R$_\text{SAD5}$ \\
\hline
Before & 12.3$\pm$1.0 & 13.6$\pm$1.2 & 14.5$\pm$1.5 & 14.2$\pm$1.9 & 14.5$\pm$2.4 & 13.0$\pm$1.9 \\
During & 12.3$\pm$3.7 & 14.3$\pm$4.5 & 15.5$\pm$4.4 & 14.7$\pm$9.0 & 14.8$\pm$1.5 & 12.8$\pm$2.0 \\
After & 12.1$\pm$1.4 & 14.5$\pm$1.4 & 13.3$\pm$1.8 & 14.3$\pm$2.4 & 13.0$\pm$1.6 & 12.8$\pm$1.4 \\ 
\hline 
\end{tabular}
\end{table*}

Because the corona is optically thin, the obtained DEM contains contribution not only from the volume occupied by the SAD but also from the rest of the column of plasma along the LOS. We found that the DEMs in the flaring region usually have a double-peak profile and the low temperature component persistently peaks at 1-2 MK. We conclude that this cooler component is mostly contributed by the foreground and background plasma in the quiet corona along the LOS, in agreement with \citet{Hannah2013}. Therefore, in this study, we utilize the aforementioned DEM$_\text{h}$ (DEM component at 4-32 MK) to analyze SADs.

Each of the six studied SADs from the two flares displays a similar pattern on the evolution of the DEM$_\text{h}$ profile, that is, a significant depression of the DEM in the temperature range of 4-32 MK, presumably due to the density depletion. This depression of DEM$_\text{h}$ suggests that at the detected altitude, SADs must have replaced most of the hot haze that would otherwise occupy the LOS volume, that is, they must have similar LOS depths to the supra-arcade haze, and furthermore they must be much rarer than the hot haze. Apparently this depression could be affected by the LOS direction, as demonstrated by the different magnitude of depression between the two events studied. In the 2014 Apr 2 event, due to the oblique viewing angle, there was more haze projected in front of or behind the SADs along the LOS, resulting in a weaker depression of DEM$_\text{h}$ than in the 2013 Oct 2 event.

Except for the depressed intensity, SADs present a negligible impact on the temperature distribution of the DEM profile. This has been further quantified using the DEM-weighted mean temperature defined conventionally as follows,
\begin{equation}
\langle T \rangle= \frac{\sum DEM(T)\times T \Delta T}{\sum DEM(T) \Delta T},
\end{equation}
For DEM$_\text{h}$ in this study, we select 4-32 MK as the sum interval to calculate $\langle T \rangle_\text{h}$ \citep{Gou2015}. Table~\ref{tbl1} lists $\langle T \rangle_\text{h}$ within all R$_\text{SAD}$s before, during, and after the passage of SADs. Generally, for each ROI, $\langle T \rangle_\text{h}$ remains almost unchanged during the investigated period. Although it seems that SAD$_{2}$ slightly increased $\langle T \rangle_\text{h}$ of R$_\text{SAD2}$ while SAD$_\text{3a}$ and SAD$_{4}$ slightly decreased $\langle T \rangle_\text{h}$ of the corresponding ROIs, these changes are within the uncertainties introduced by the DEM. \citet{Hanneman2014}, using \verb+xrt_dem_iterative2.pro+ in SSW, suggested that SADs are cooler than the surrounding plasma but hotter than the coronal background, based on the peak temperature of the DEM profile. We would like to point out that most likely, with the LOS integration effect, neither the peak temperature nor $\langle T \rangle_\text{h}$ represent the temperature of SAD itself, but the temperature of the remaining haze along the LOS.

In comparison with numerical simulations, we found no indication of enhancement of DEM$_\text{h}$ ahead of the SADs in either flare studied, in agreement with the observations by \citet{Innes2014}. Arguably one cannot completely exclude the possible presence of a high-emission shock ahead, which could be too thin to be seen by AIA, although the shocked region is comparable in size as the void in \citet{Scott2013}. We found that the SAD-associated depression of DEM$_\text{h}$ can be recovered within a few minutes. In the 2014 April 2 flare, there are occasions that SADs apparently fall sequentially through the same route, but it is possible that different routes are aligned along the LOS which is oblique to the axis of the flare arcade. Therefore, it may not be necessary to resort to continuous reconnections to keep SADs from being filled from behind as in \citet{Cassak2013}, if they are inherently discrete features. 

As far as plasma properties are concerned, the simulated SADs in \citet{Guo2014} and \citet{Cecere2015} have similar characteristics, which are of low density ($\sfrac{1}{3}-\sfrac{1}{2}$ of the surroundings) but high temperature (more than 20 MK). However, EM of the simulated SADs is only one order of magnitude smaller than the surroundings, and should therefore manifest as a bump in the tail of the DEM profile, resulting in an increase in $\langle T \rangle_\text{h}$. In contrast, our results show that DEM$_\text{h}$ is depressed by more than one order of magnitude but that $\langle T \rangle_\text{h}$ remains more or less constant with the presence of SADs. For instance, EM$_\text{h}$ (integrating DEM$_\text{h}$ over the temperature) is reduced from $10^{27}$ to $10^{25}$ (cm$^{-5}$) with the passage of SAD$_{1}$. However, one must keep in mind that AIA is insensitive to hot plasma with temperature above 20 MK, so whether SADs have extremely hot temperatures remains an open question. On the other hand, the high temperature helps prevent the simulated low-density SADs from being filled by surrounding plasma, but may not be necessary if SADs are magnetic entities with enhanced magnetic pressure \citep{Liu2013}.

Thus, our study supports this idea that SADs are indeed devoid of plasma, with negligible DEM. By monitoring the change of DEM with the passage of SADs, we demonstrate that SADs are able to squeeze out the hot and dense haze, meaning that the hot component of DEM ($>4$ MK) is reduced roughly to the coronal background level, without significant change of the profile. It is not surprising that the observations have discrepancies with recent models, qualitatively or quantitatively, considering that the simulations seldom take into account thermal conduction and radiative losses, and that the assumption underlying the DEM method, local thermal equilibrium, may not be valid in the reconnection outflow where SADs are supposedly formed. Further efforts are needed from both observational and theoretical aspects to improve our understanding of SADs.

%-----------------------------------------------------------------
\begin{acknowledgements}
The authors thank Dr.~Iain Hannah for help with the DEM inversion code. The authors are grateful to the team of SDO for making the data and processing code freely available. This research was supported by NSF under grants AGS 1153226, 1348513 and 1408703. R.L. acknowledges the Thousand Young Talents Program of China and NSFC 41474151.
\end{acknowledgements}

\begin{appendix}

\section{Running difference movies}

\begin{figure}
\centering
\includegraphics[width=8cm]{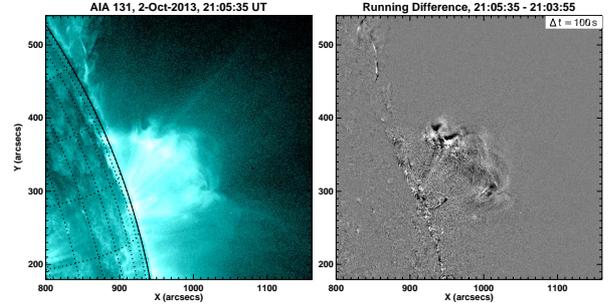}
\caption{One frame of the movie1, available online, showing AIA 131 \r{A} and corresponding running difference images for 2013 Oct 02 event. The varying $\Delta t$ is due to bad frames caused by the HMI calibration.}
\label{figA1}
\end{figure}

\begin{figure}
\centering
\includegraphics[width=8cm]{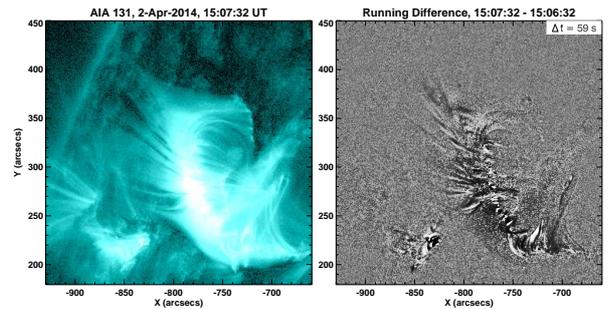}
\caption{One frame of the movie2, available online, showing AIA 131 \r{A} and corresponding running difference images (with enhanced visibility) for 2014 Apr 02 event.}
\label{figA2}
\end{figure}

\end{appendix}


\bibliographystyle{aa}
\bibliography{manuscript}

\end{document}